\documentclass[showpacs,preprintnumbers,amsmath,amssymb,superscriptaddress,floatfix,twocolumn]{revtex4-1}

\usepackage{graphicx}
\usepackage{amsfonts}
\usepackage{ulem}
\usepackage{color}
\usepackage{url}
\input{colordvi.tex}

\begin{document}

\title{Test for anisotropy in the mean of the CMB temperature
fluctuation in spherical harmonic space}

\author{Daichi Kashino}
\author{Kiyotomo Ichiki}
\author{Tsutomu T. Takeuchi}

\affiliation{%
Department of Physics and Astrophysics, Nagoya University, Nagoya
464-8602, Japan
}

\email{daichi@nagoya-u.jp}
%\email{kashino.daichi@b.nagoya-u.jp}

%\affiliation{%
%place2
%}

%\author{author3}
%\affiliation{%
%place3
%}

%\date{\today}
\begin{abstract}

The standard models of inflation predict statistically homogeneous and
isotropic primordial fluctuations, which should be tested by
observations.
In this paper we illustrate a method to test the statistical isotropy of the mean of the cosmic microwave background temperature fluctuations
in the spherical harmonic space and apply the method to the Wilkinson Microwave Anisotropy Probe seven-year observation data.
A classical method to test a mean, like the simple
Student's t test, is not appropriate for this purpose because the
Wilkinson Microwave Anisotropy Probe data contain anisotropic instrumental noise and suffer from the effect of the mask
for the foreground emissions which breaks the statistical independence.
Here we perform a band-power analysis with Monte Carlo simulations in which  we take into account the anisotropic noise and the mask.
We find evidence of a non-zero mean at 99.93 \% confidence level in a
particular range of multipoles.
The evidence against the zero-mean
assumption as a whole is still significant at the 99 \% confidence
level even if the fact is taken into account that we
have tested multiple ranges.

\end{abstract}
\maketitle

%%% SECTION 1 %%%
\section{introduction}
Inflation provides a successful mechanism of generating primordial
density perturbations that give rise to the large-scale structure
of the Universe and temperature anisotropy in the cosmic microwave
background (CMB).
One of the important consequences of inflation
models, which should be tested with observations, is that they predict
statistically homogeneous and isotropic Gaussian fluctuations with
nearly scale invariant power spectrum \cite{2005ppci.conf..235L}.

So far, tests of the scale invariance
and/or the Gaussianity of primordial fluctuations 
have been done intensively. 
In those tests, the statistical homogeneity and isotropy are often implicitly
assumed.  This assumption, however, should be verified with observational test.
For example, the statistical homogeneity of the large-scale structure
was tested by comparing the means of the
galaxy distributions in different directions
\cite{2011CQGra..28p4003S}.  
For the CMB temperature field,
the anisotropy of the power spectrum has rigorously been tested,
particularly in the context of the hemispherical asymmetry
\cite{ 2004MNRAS.349..313P,
2004ApJ...605...14E, 2004ApJ...609.1198E, 2004MNRAS.354..641H,
2008PhRvD..78f3531B, 2009ApJ...704.1448H, 2009ApJ...699..985H}. 
Similar tests in real space of the mean,
variance, skewness and kurtosis were performed in
\cite{2008JCAP...08..017L}.

In this paper we illustrate a method to test the statistical isotropy of the mean of the CMB temperature fluctuations 
in the spherical harmonic space, 
and we apply it to the Wilkinson Microwave Anisotropic Probe (WMAP) seven-year observation data.
This method can be potentially useful
for the forthcoming PLANCK and future CMB surveys,
for which contaminations of the CMB maps by  instrumental noises
will take place for higher multipole range.

The CMB temperature field can be expressed as
a background value $T_0$ with associated fluctuations $\Delta T$,
\begin{equation}
T (\hat{n}) = T_0 (\hat{n}) + \Delta T (\hat{n})~,
\end{equation}
where $\hat{n}$ is a unit direction vector. 
Under a null hypothesis that the background $T_0$ is isotropic, $T_0
(\hat{n}) = T_0$ and the fluctuations $\Delta T(\hat{n})$ defined over
the full sky can be expanded in terms of spherical harmonics as
\begin{equation}
\Delta T ( \hat{n}) =
\sum^{\infty}_{\ell =1}\sum ^{\ell}_{m=-\ell} a_{\ell m} Y_{\ell m}(\hat{n})~,
\end{equation}
with
\begin{equation}
a_{\ell m}=
\int d\Omega ( \hat{n}) \Delta T (\hat{n}) Y^\ast _{\ell m} (\hat{n})~,
\end{equation}
where $Y_{\ell m}(\hat{n})$ are the spherical harmonic functions
evaluated in the direction $\hat{n}$. 
If the primordial fluctuations are statistically homogeneous and
isotropic in the mean and if they obey the Gaussian distribution,
$2\ell+1$ spherical harmonic
coefficients $a_{\ell m}$s for each $\ell$ are
independent Gaussian variables with the following properties
\begin{equation}
\left< a_{\ell m} \right> =0~,
\label{eq:alm0}
\end{equation}
and
\begin{equation}
\left< a_{\ell m} a_{\ell ^\prime m^\prime }^\ast \right> = \delta _{\ell \ell^\prime} \delta_{mm^\prime}C_{\ell}~,
\label{eq:cl}
\end{equation}
where $\left< \cdot \cdot \cdot \right>$ denotes the ensemble average,
$C_\ell$ is the ensemble average power spectrum and $\delta $ is the
Kronecker symbol.

This null hypothesis that the means of $a_{\ell m}$s are zero
[Eq. (\ref{eq:alm0})] should be verified with observation.
Our aim of this paper is to test this null hypothesis against the alternative
hypothesis that the means of $a_{\ell m}$s are non-zero. 
One of the difficulties in this test is
to decorrelate the neighboring modes in harmonic space 
whose correlations are caused by a cut sky mask. 
In order to overcome this problem,
we use the band-power decorrelation analysis
following the method in \cite{2008PhRvD..78l3002N}. 
The method is to decompose the correlated spectrum
to mutually independent band-power spectrum 
by diagonalizing the covariance matrix.
We run Monte Carlo simulations
to calculate the covariance matrix of the mean distribution,
including anisotropic instrumental noise and the effect of the mask.

Recently,
Armendariz-Picon tested this zero-mean hypothesis
in a smart analytic way with a symmetric cut sky mask
to obtain uncorrelated and independent variables,
with the noise contributions neglected \cite{2011JCAP...03..048A}. 
The author found significant evidence for
nonzero means of $a_{\ell m}$s in a particular multipole range.
He concluded, however,
that this evidence is statistically insignificant because
the signal is high only at one bin among eight different multipole bins.
The main difference between our analysis and his is that we make extensive use of
Monte Carlo simulations to keep the cosmological information as much as
possible while taking into account the noise contributions and the effect of
the WMAP mask.

This paper is organized as follows.
In Sec. II we briefly review the main properties of the CMB temperature field,
its spherical harmonic analysis and the cut sky mask.
We describe the method of our analysis based on
Monte Carlo simulations with the WMAP seven-year data in Sec. III.
Our results are presented in Sec. IV and we provide a discussion
in Sec. V.
We conclude our findings in Sec. VI.

%%%  SECTION 2 %%%
\section{Temperature field and Mask}
The temperature field $\Delta T_\mathrm{obs}$ observed
with instruments involves a convolution with the detector beam window
$B$ and the detector noise $\Delta T_\mathrm{noise}$.
In addition, since we have to divide the all sky into finite pixels,
we introduce a pixel smoothing kernel $K$,
and write the temperature field as
\begin{equation}
\Delta T_\mathrm{obs}  = K \ast \left[ B \ast \Delta T_\mathrm{real} + \Delta T_\mathrm{noise} \right]~,
\end{equation}
%\begin{equation}
%\label{eq:DeltaT}
%\Delta T_\mathrm{obs}\left( \hat{n} \right)=\int d\Omega _{\hat{n}^\prime}\Delta T_\mathrm{real}\left( \hat{n}^\prime \right) w_{\rm{b}} \left( \hat{n}-\hat{n}^\prime \right) +\Delta T_\mathrm{noise} \left( \hat{n} \right),
%\end{equation}
where the star ``$\ast$" denotes the convolution and 
$\Delta T_\mathrm{real}$ is a superposition of CMB temperature field $\Delta_\mathrm{CMB}$ and foreground emission $F$;
\begin{equation}
\Delta T_\mathrm{real}(\hat{n})=\Delta T_\mathrm{CMB}(\hat{n})+F(\hat{n}) .
\end{equation}
The foregrounds consist of, for example,
the dust emission and synchrotron
radiation from our own Galaxy and emissions from extragalactic objects.
The WMAP team provides the sky map data with the foregrounds reduced by using
appropriate foreground templates \cite{2011ApJS..192...15G}.
However, the cleaning procedure does not completely remove the foreground contamination
along the galactic disk and that from extragalactic objects.
To avoid the effect of the residual foreground, the contaminated regions have to be masked out.
The mask is defined by a position-dependent weight function $M (\hat{n})$ as
\begin{eqnarray}
M \left(\hat{n} \right)=\left\{ \begin{array}{ll}
0 & \qquad \mbox{for contamination region} \\
1 & \qquad \mbox{otherwise}. \\
\end{array} \right.
\end{eqnarray}
By definition, $M\left( \hat{n} \right) \left[ K \ast B\ast F \right] \left( \hat{n} \right) =0$.
Then, by construction, we get the masked sky $\Delta T_M $ which does not include foregrounds,  
\begin{equation}
\Delta T_\mathrm{M}   \left( \hat{n} \right) = M \left( \hat{n} \right) \left[ K\ast B\ast \Delta T_\mathrm{CMB} +K\ast \Delta T_\mathrm{noise}\right] \left( \hat{n} \right).
\label{eq:DeltaT_M}
\end{equation}

Equations (\ref{eq:alm0}) and (\ref{eq:cl}) state that the $2\ell +1$
variables $a_{\ell m}$ form a set of normally distributed independent
variables with a variance $C_\ell$.
However, in reality, we cannot 
obtain the {\it true} $a_{\ell m}$ but only the {\it pseudo} $a_{\ell m}$
($a^\mathrm{mask}_{\ell m}$) from the masked sky map.  Rewriting
Eq. (\ref{eq:DeltaT_M}) in harmonic space,
a $a^\mathrm{mask}_{\ell m}$ from the masked sky map is given by
\begin{equation}
 a^\mathrm{mask}_{\ell m} = \sum_{\ell^\prime m^\prime} M_{\ell m, \, \ell^\prime m^\prime} a^\mathrm{all\, sky}_{\ell m}~,
\end{equation}
where $a^{\mathrm{all\, sky}}_{\ell m}$ is from the all sky CMB map
which we can never obtain and $M_{\ell m, \, \ell^\prime m^\prime}$ is
the convolution matrix of the mask.
The detector pixel noise is
well described by a Gaussian distribution with zero mean
as shown in the WMAP papers \cite{2003ApJS..148...29J, 2007ApJS..170..263J}.
Therefore if ensemble averages of {\it true} $a_{\ell m}$s are zero,
$\left< a_{\ell m}^\mathrm{all \, sky }\right> =0$, the ensemble
averages of {\it pseudo} $a_{\ell m}$s are zero,
$\left< {a^\mathrm{mask}_{\ell m}} \right> = 0$.
The details of the convolution matrix are not essential in the present analysis,
but the matter of importance is that $a_{\ell m}$s have correlations
with neighboring modes.

%%%  SECTION 3 %%%
\section{Data and Analysis}

\subsection{Data} 

For the main analysis in this paper, we use the WMAP seven-year
foreground-reduced sky maps \cite{2011ApJS..192...14J} from differential
assemblies (DAs) Q1, Q2, V1, V2, and W1, W2, W3, W4, pixelized in the
HEALPix \cite{2005ApJ...622..759G} sphere pixelization scheme with
the resolution parameter $N_\mathrm{side} = 512$.
These sky maps are provided
at the LAMBDA web site \cite{LAMBDAwebsite}.
We co-added them using inverse noise pixel weighting and form either
individual frequency combined maps (Q, V, W) or an overall combined map
(Q+V+W) to increase the signal to noise ratio according to
\begin{equation}
T(i)=\frac{1}{W(i)}\sum ^{j_e}_{j=j_s} w_j(i)T_j(i)~,
\label{eq:combine}
\end{equation}
where $i$ and $j$ are the pixel and
DA indices, respectively. The noise weighting function $w_j(i)$ is given
by $w_j(i) = N^{\rm obs}_{j} (i)/ \sigma^2_{0j}$, where $\sigma_{0j}$ is
the noise root-mean square for the $j$th DA, and $N^{\mathrm{obs}}_j(i)$
is the number of observations of the $i$th pixel for the $j$th DA, and
$W(i) = \sum^{j_e}_{j_s} w_j(i)$.

For the mask, to eliminate foreground contamination, we use the extended
temperature analysis mask (KQ75y7 sky mask) \cite{2011ApJS..192...14J}.

We use a power spectrum $C_{\ell}$ file and a beam transfer function
file to produce realizations of the temperature map.
The power spectrum $C_{\ell}$ we use is the best fitting $C_{\ell}$
to the WMAP seven-year data set \cite{2011ApJS..192...16L}
in the $\Lambda$CDM (+sz+lens) model.
The $C_\ell$ and the beam transfer function of each DA
are also available at the LAMBDA web site.

Our conclusions are based on the results with the overall combined map (Q+V+W).
However, we analyze also the individual frequency maps (Q, V, W)
to examine effects of the residual foreground emissions
and the anisotropic instrumental noise.
Because their properties are different for frequency bands.

\subsection{Monte Carlo simulation}

We carry out Monte Carlo simulations to test
whether the temperature map observed by the WMAP has indeed zero mean in
harmonic space.  Our analysis consists of four steps.
First, we prepare 10 000 realizations of the sky map
from the normally distributed
$a_{\ell m}$s generated from the same underlying power spectrum.
Second, we apply the cut sky mask to these Gaussian maps and the
temperature maps obtained by the WMAP, and extract $a^\mathrm{mask}_{\ell
m}$s from each map through the harmonic analysis.
Third, for the set of $a^\mathrm{mask}_{\ell m}$s obtained in the previous step,
we calculated the means of $a^{\mathrm{mask}}_{\ell m}$s
for each multipole moment $\ell$.
Finally, we decompose the correlated spectrum of the mean to
the independent band spectrum and calculate $p$ values
of the WMAP observed values by
comparing with the distributions realized by simulations.

The method of generating Gaussian sky maps consists of three
steps described in detail below.
There are two important points in generating Gaussian sky maps.
The first is that Gaussian maps should be
generated using the same power spectrum as the WMAP sky map.
The second is that we must add random noise taking into account the inhomogeneous
nature of $N_\mathrm{obs}$ to each Gaussian sky map.

{\it Step 1 : Generate the pure Gaussian sky maps}

We prepared 10 000 Gaussian sky maps by HEALPix IDL facility {\it
isynfast} for each DA, which are generated from standard
normally distributed random $a_{\ell m}$s from a given power spectrum
$C_{\ell}$.
The $C_\ell$ inputted is the best-fit ($\Lambda$CDM
(+sz+lens) model ) of the WMAP seven-year data set.
The resolution of the maps is $N_\mathrm{side} = 512$
which is the same as the resolution of WMAP data we use.
We use the same random seed array for all DAs.
That is, we generate random sample of 10 000 different universes.

{\it Step 2 : Add noise}

According to the WMAP team \cite{2003ApJS..148...29J, 2007ApJS..170..263J},
the instrumental noise is expected to exhibit a Gaussian distribution
and the noise variance in a given pixel is
inversely proportional to the number of observations of that pixel,
$N_\mathrm{obs}$.
The rms noise per observation $\sigma_0$ for each DA is given by the WMAP team.
Since a Gaussian map generated by {\it isynfast} contains no noise, 
the noise should be added on it.
We add random noise considering the inhomogeneity of $N_\mathrm{obs}$ to all
pure Gaussian maps.

{\it Step 3 : Combine Gaussian maps}

We combine Gaussian maps whose seed number is the same using
Eq. (\ref{eq:combine}) and form either individual frequency maps (Q, V,
W) or an overall combined map (Q+V+W).

\subsection{Test Statistic}

Throughout this analysis we deal with {\it real} coefficients.
The real coefficients $a_{\ell m}$s and conventional complex coefficients
$\mathcal{A}_{\ell m}$s are related to each other by
\begin{equation}
a_{\ell m} = \left\{
\begin{array}{lll}
-\sqrt{2} \, \mathrm{Im} \, \mathcal{A}_{\ell -m} &\quad & \mbox{if   } m<0 \nonumber \\
\mathcal{A}_{\ell m} &\quad & \mbox{if   } m=0 \nonumber \\
\sqrt{2} \, \mathrm{Re} \, \mathcal{A}_{\ell m} &\quad & \mbox{if  } m>0.
\end{array}
\right.
\end{equation}

We define the mean spectrum as 
\begin{equation}
M_\ell = \frac{\sum_{m = -\ell}^\ell a^\mathrm{mask}_{\ell m} / \sqrt{\mathrm{Var}_\ell}}{\sqrt{2\ell +1}}~.
\label{eq:M_ell}
\end{equation}
Here $\mathrm{Var}_\ell$ is the average of the
variance of $a_{\ell m}$s calculated from the 10 000 realizations,
\begin{equation}
\mathrm{Var}_\ell = \frac{1}{\mathcal{N}}\sum_{\alpha = 1}^{\mathcal{N}} \frac{\sum_m a^{\mathrm{mask},\alpha}_{\ell m}}{2\ell +1}~,
\label{eq:Var_ell}
\end{equation}
where $a^{\mathrm{mask},\alpha}_{\ell m}$ represents the multipole
coefficient of the $\alpha$th realization and $\mathcal{N}=10 000$ is
the total number of realizations.  This mean spectrum is normalized by
the number of samples in a given $\ell$, namely, $2\ell +1$.

The neighboring $\ell$ modes of the mean spectra are strongly correlated
with each other, and the number of independent modes is much
smaller.
To obtain statistically independent variables, we define the
binned mean spectrum $\mathcal{M}_i$ as,
\begin{equation}
\mathcal{M}_i  = \frac{\sum_{\ell=\ell_{-,i}}^{\ell_{+,i}} \sum_{m=-\ell}^{\ell} a^\mathrm{mask}_{\ell m} / \sqrt{\mathrm{Var}_\ell} }{\sqrt{\sum_{\ell =\ell_{-,i}}^{\ell_{+,i}} (2\ell+1)}}~,
\label{eq:binned-mean-spec}
\end{equation}
where $i$ is the index of the multipole bin from $\ell = \ell_{-,i}$
to $\ell= \ell_{+,i}$.  This binned statistical variable is normalized
by the sample number in a bin, i.e., $\sum_{\ell = \ell_{-,i}}^{\ell_{+,i}}
(2\ell +1)$.

The covariance matrix of $\mathcal{M}_i$ is defined by 
\begin{eqnarray}
K_{ij} \equiv \frac{1}{\mathcal{N}} \sum_{\alpha = 1}^{\mathcal{N}} \mathcal{M}^\alpha_i \mathcal{M}^\alpha_j~,
\label{eq:K}
\end{eqnarray}
where $\mathcal{M}_i^\alpha$ represents the value of the $i$th multipole bin in the $\alpha$th realization.
$K$ is a real symmetric matrix with its dimension equal to the number of multipole bins, $N$, 
and it can be diagonalized by a real unitary matrix $U$,
\begin{equation}
UKU^\dagger = \mbox{diag} ( \lambda_1,\, \lambda_2, ...,\, \lambda_N ) \equiv \Lambda ~,
\end{equation}
where $\lambda_i$'s the eigenvalues of $K$.
A window matrix $W$ is defined as \cite{2008PhRvD..78l3002N}
\begin{equation}
W_{ij} = \frac{(K^{-1/2})_{ij}}{\sum^{N}_{m=1} (K^{-1/2})_{im}}~,
\label{eq:W}
\end{equation}
where
\begin{equation}
K^{-1/2} \equiv U^\dagger \Lambda^{-1/2} U~,
\end{equation}
and
\begin{equation}
\Lambda ^{-1/2} \equiv \mbox{diag} ( \lambda_1^{-1/2},\, \lambda_2^{-1/2}, ...,\, \lambda_N^{-1/2} )~.
\end{equation}
We define a new decorrelated statistical variable $S^\alpha _i$ by using
the window matrix as
\begin{equation}
S_i^\alpha \equiv \sum^{N}_{j=1} W_{ij} \mathcal{M}^\alpha_j~.
\end{equation}
Then the correlation matrix of the variable is diagonal and reads
\begin{eqnarray}
\frac{1}{\mathcal{N}} \sum^{\mathcal{N}}_{\alpha = 1} S^\alpha_i
 S^{\alpha}_j &=& \left( W K ^t W \right)_{ij} 
\nonumber \\ &=& \left[ \sum_{m = 1}^{N} \left( K^{-1/2} \right) _{im}\right] ^{-2} \delta_{ij}~,
\label{eq:S_i^a}
\end{eqnarray}
where $^t W$ denotes the transposed matrix.

%%%  SECTION 4 %%%
\section{Result}
\begin{figure}[htbp] %  figure placement: here, top, bottom, or page
   \centering
   \includegraphics[width=\linewidth]{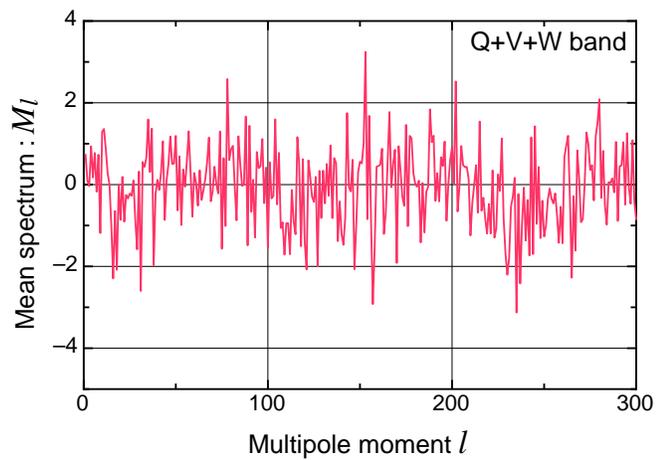} 
   \caption{The mean spectrum, $M_\ell$, of the overall combined map (Q+V+W), defined in Eq. (\ref{eq:M_ell}).  }
   \label{fig:mean_sp}
\end{figure}

\begin{figure}[htbp] %  figure placement: here, top, bottom, or page
%   \centering
\hspace*{-2cm} \includegraphics[width=\linewidth,
   angle=-90]{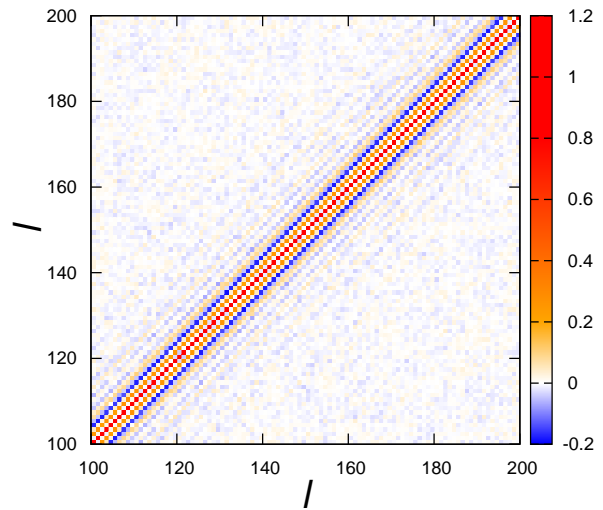} \caption{The covariance
   matrix of the mean spectrum, $M_\ell$, calculated by Monte Carlo
   simulations. The central red strip is diagonal element array and the
   surrounding blue strips indicate negative correlation. Only a part
   ($100 \leq \ell \leq 200$) of the covariance matrix whose dimension
   is 300 is shown here.}  \label{fig:covmat}
\end{figure}

We calculate the mean spectrum, $M_\ell$, over the multipole range from $\ell
= 1$ to $\ell=300$ of the WMAP seven-year data, which is shown in
Fig. \ref{fig:mean_sp}.
Figure \ref{fig:covmat} shows the covariance matrix of $M_\ell$
calculated for 10 000 realizations made by Monte Carlo simulations.
We can see that neighboring $\ell$ modes are correlated with each other
up to the range $\Delta \ell \sim 10$.

Therefore we divide multipole range from $\ell=1$-$300$ 
into the bins whose size is $\Delta \ell =6$,
$10$, or $20$ to calculate the binned mean spectrum,
$\mathcal{M}_i$. 
We calculate the covariance matrix $K_{ij}$
defined by Eq. (\ref{eq:K}) and the window matrix $W_{ij}$ defined by
Eq. (\ref{eq:W}).  Then we construct decorrelated statistical variables
of the WMAP seven-year data using $W_{ij}$.

\begin{figure}[htbp] %  figure placement: here, top, bottom, or page
   \centering \includegraphics[width =
   \linewidth]{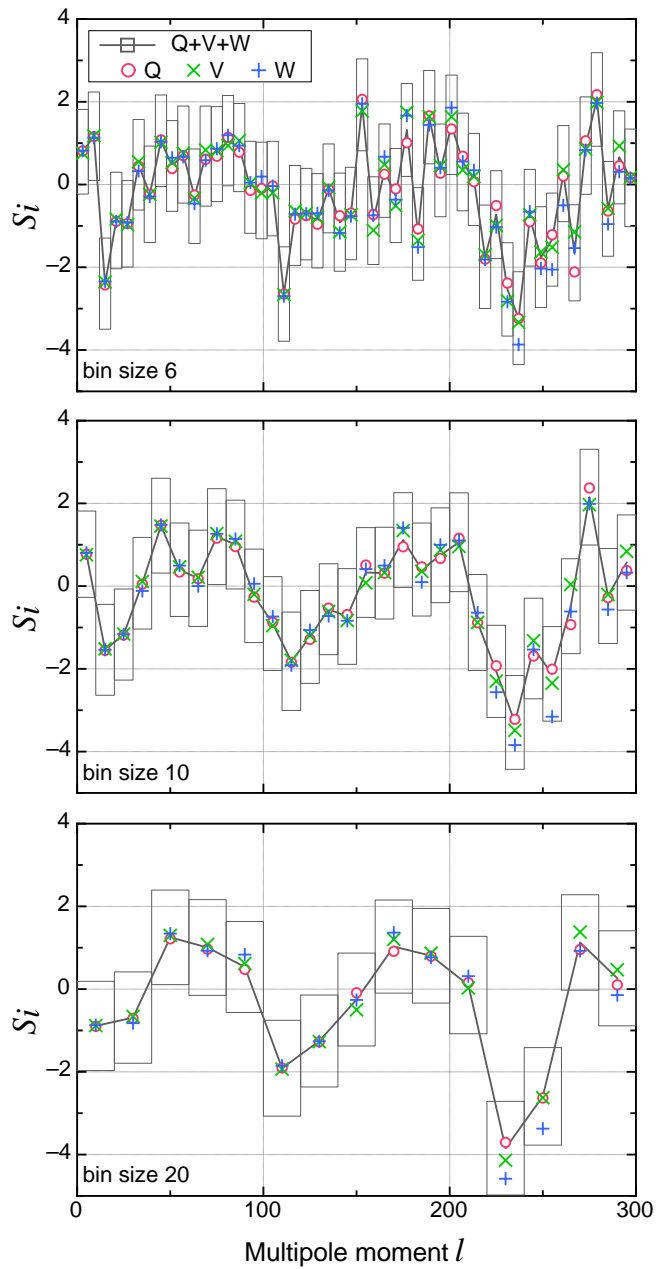} 
   \caption{The decorrelated band mean spectrum obtained by the seven-year WMAP data.
   The boxes and solid line are the result of the overall combined map
   (Q+V+W) and color data points are the results of the individual
   frequency maps (Q, V, W).  The horizontal width of boxes indicates
   the bin size $\Delta \ell = 6$ (top panel), 10 (middle panel) and 20 (lower panel) 
   for calculating the binned mean
   spectrum, $\mathcal{M}_i$.}  \label{fig:band-mean-spec}
\end{figure}

The decorrelated band mean spectra of the WMAP seven-year data are
represented in Fig. \ref{fig:band-mean-spec}.
In these panels, the solid line indicates the
results from the combined map (Q+V+W),
\begin{equation}
S_i = \sum^{N}_{j=1} W_{ij} \mathcal{M}_j~.
\end{equation}
The box represents the standard deviation for each multipole bin,
\begin{equation}
\sigma_i = \left[ \frac{1}{\mathcal{N}} \sum^{\mathcal{N}}_{\alpha = 1} \left( S^\alpha_i \right)^2 \right] = \left[ \sum_{m = 1}^{N} \left( K^{-1/2} \right) _{im}\right] ^{-1}~,
\end{equation}
and the bin size are taken as $\Delta \ell = 6$, $10$ or $20$ in
calculating the binned mean spectrum, $\mathcal{M}_i$.
The color data
points show the results from the individual frequency maps (Q, V, W).

\begin{figure}[htbp] %  figure placement: here, top, bottom, or page
   \centering
   \includegraphics[width = \linewidth]{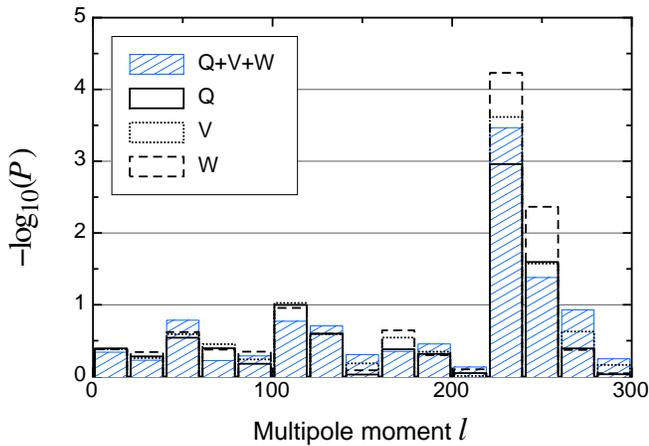} 
   \caption{The $p$ values of the data points in
 the lower panel in Fig. \ref{fig:band-mean-spec}, for the bin size $\Delta \ell = 20$.}
   \label{fig:p-value}
\end{figure}

We can see, in the top panel and the middle panel in Fig. \ref{fig:band-mean-spec},
that the data points in the multipole
range $\ell \sim 210$-$260$ are localized in the
negative side as a cluster despite the fact that these data points are
random and mutually independent.
This negative localization is clearly evident in the lower panel in Fig. \ref{fig:band-mean-spec}
where the bin width is taken as $\Delta \ell = 20$. 
It is particularly worth noting that 12th data point
(on the multipole bin $\ell =221$-$240$)
indicates the nonzero mean evidence with 99.93\% confidence
level [C.L.] ($\alpha$ ($p$ value) $=$ 0.068\%) for the combined map (Q+V+W).
Figure \ref{fig:p-value} and Table \ref{tb:p-value} show the $p$ values of the data points
in the lower panel in Fig. \ref{fig:band-mean-spec}.
Clearly, the significance is above at
the 99\% confidence level even for the individual Q, V, W maps, and in fact,
for the V, W maps the significance becomes even larger above the 99.9\% C.L.

%%%  SECTION 5 %%%
\section{Discussion}
Since in the WMAP observation data the instrumental noise becomes dominant
at higher multipoles (see Appendix A), 
We have used the multipole range up to $\ell = 300$.
The method we have described can be useful for the forthcoming PLANCKs
and future CMB surveys , for which contaminations of the CMB maps
by the instrumental noises will take place for higher multipole range.

We can see that the statistics in the above-mentioned
figures exhibit the same behavior across the three frequency bands (Q,
V, W).  Therefore it does not seem that these results are due to
residual foregrounds and/or detector noise characteristics which depends
on the frequency band, but the cosmological CMB signals.

\begin{figure}[htbp] %  figure placement: here, top, bottom, or page
   \centering
   \includegraphics[width=\linewidth]{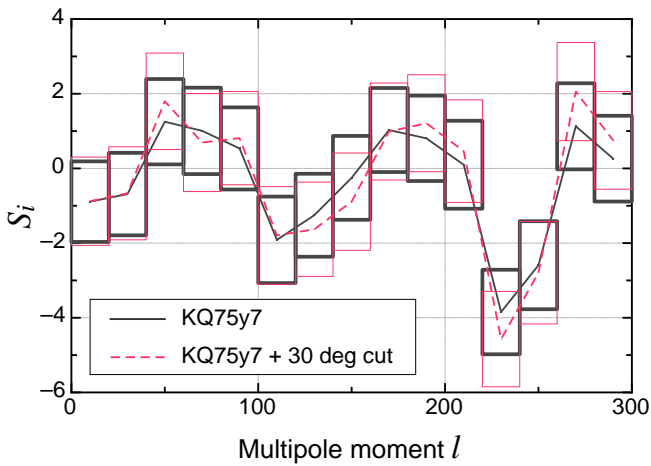}
   \caption{The decorrelated band mean spectra for two different
   masks. The black solid line (thick boxes) is the result for the conventional mask (KQ75y7) and the red
   dashed line (thin boxes) is for the more expansive mask (KQ75y7 + cut $|b| \leq 30^\circ$).}
   \label{fig:30deg_datapoint}
\end{figure}

\begin{figure}[htbp] %  figure placement: here, top, bottom, or page
   \centering
   \includegraphics[width = \linewidth]{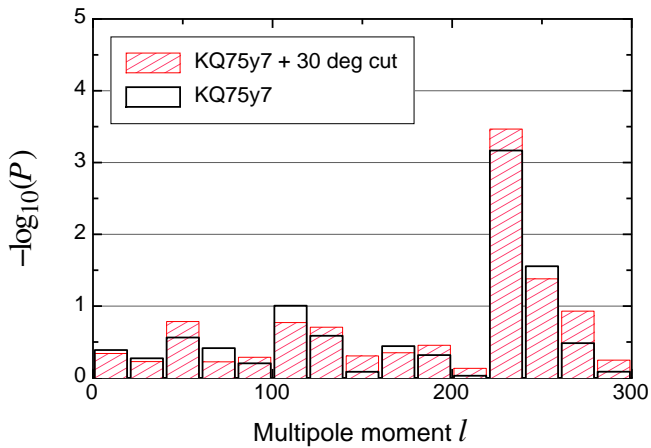} 
   \caption{The $p$ values of data points in Fig. \ref{fig:30deg_datapoint}.}
   \label{fig:30deg_pvalue}
\end{figure}

To eliminate residual foreground emissions more
conservatively and confirm these results, we repeat
the same analysis with a more extensive mask which cuts the
region in galactic latitude $|b| \leq 30^\circ$ in addition to KQ75y7
sky mask \cite{2011ApJS..192...14J}.
Again we obtain consistent results,
and in fact, the significance becomes even larger
against the zero mean hypothesis as shown in
Figs. \ref{fig:30deg_datapoint} and \ref{fig:30deg_pvalue}.
The $p$ values are summarized in Table \ref{tb:p-value}.
The $p$ value of the 12th multipole bin $\ell = 221$-$240$
goes down to $\alpha = 0.034\%$ from 0.068\%,
albeit the standard deviation $\sigma$ becomes larger
due to the loss of the sky area to analyze.

Among these analyses of three bin sizes, the results of the bin size
$\Delta \ell = 6$ keep the original information of the CMB signal as
much as possible and represent the finest structure of the means of
$a_{\ell m}$s.  However, we find that the most intriguing results are
for $\Delta \ell = 20$ because they represent the negative localization
of the means in the multipole range $\ell = 221$-$240$ most significantly.
  
\begin{table}[htbp]
\caption{The $p$ values for each CMB map with the KQ75y7 mask (Q, V, W and Q+V+W) and the extensive mask(KQ75y7 + cut $|b| \leq 30^\circ$) (Q+V+W only). The bin column indicates the multipole bounds for calculation the binned mean spectrum, $\mathcal{M}_i$.
The $p$ value is anomalously small in 12th multipole bin $\ell = 221$-$240$.}
\begin{center}
\begin{tabular}{lcccccccc}
\hline
\hline
\multicolumn{2}{c}{bin}  &  & \multicolumn{4}{c}{KQ75y7 mask}  &  &  + $30^\circ$ cut \\
$\ell_\mathrm{min}$ & $\ell_\mathrm{max}$ & & Q & V & W & Q+V+W &  & Q+V+W \\
\hline
1 & 20  & & 40.5\% & 41.4\% & 41.7\% & 40.9\% &  & 45.7\% \\
21 & 40  & & 52.4\% & 55.1\% & 45.5\% & 53.3\% &  & 59.3\% \\
41 & 60  & & 28.7\% & 25.6\% & 24.0\% & 27.4\% &  & 16.4\% \\
61 & 80  & & 40.3\% & 35.3\% & 42.1\% & 38.5\% &  & 59.7\% \\
81 & 100  & & 66.3\% & 57.6\% & 45.0\% & 62.7\% &  & 51.6\% \\
101 & 120 & & 10.0\% & 9.44\% & 11.1\% & 9.87\% &  & 17.0\% \\
121 & 140 & & 25.3\% & 25.5\% & 25.3\% & 25.9\% &  & 19.7\% \\
141 & 160 & & 93.7\% & 65.6\% & 81.6\% & 82.2\% &  & 49.4\% \\
161 & 180 & & 41.6\% & 28.6\% & 22.7\% & 36.1\% &  & 44.7\% \\
181 & 200 & & 48.4\% & 45.0\% & 50.3\% & 48.1\% &  & 35.1\% \\
201 & 220 & & 89.4\% & 98.0\% & 79.2\% & 93.3\% &  & 73.6\% \\
{\bf 221} & {\bf 240} & & {\bf 0.110}\% & {\bf 0.024}\% & {\bf 0.006}\% & {\bf 0.068}\% &  & {\bf 0.034}\% \\
241 & 260 & & 2.53\% & 2.66\% & 0.431\% & 2.78\% &  & 4.20\% \\
261 & 280 & & 40.5\% & 23.6\% & 42.5\% & 32.8\% &  & 11.8\% \\
281 & 300 & & 92.8\% & 68.8\% & 90.0\% & 82.0\% &  & 56.6\% \\
\hline
\hline
\end{tabular}
\end{center}
\label{tb:p-value}
\end{table}

On ground that we have tested $n = 15$ independent bins, we should
evaluate significance as a whole to draw a
conclusion against the null hypothesis.  If we would like to advocate
the anomalousness of the 12th multipole bin in the lower panel in
Fig. \ref{fig:band-mean-spec} at the significant level
$\alpha_\mathrm{tot}$, we should require the
$\alpha$ of each individual test to satisfy
\begin{equation}
1-\alpha_\mathrm{tot} = (1-\alpha )^{n}~.
\end{equation}
For $n=15$ independent bins and $\alpha_\mathrm{tot} = 5\%$, this yields $\alpha =
0.34\%$, and for $\alpha_\mathrm{tot} = 1\%$, $\alpha = 0.067\%$. 
Table \ref{tb:alpha_tot} represents the values of $\alpha$ of the 12th multipole
bin and the corresponding values of $1-\alpha_\mathrm{tot}$ for the combined
map (Q+V+W) and the individual Q, V, W maps.
Clearly, this evidence against the zero-mean hypothesis is still significant,
keeping the 99\% C.L. (Q+V+W).

\begin{table}[htbp]
\caption{The values of $\alpha$ of the 12th bin and the corresponding values of $1-\alpha_\mathrm{tot}$.}
\begin{center}
\begin{tabular}{lrrcc}
\hline
\hline
 & \multicolumn{2}{c}{KQ75y7} & \multicolumn{2}{c}{KQ75y7 + 30 deg cut}\\
\makebox[1cm][c]{} & 
\makebox[1.5 cm][c]{$\alpha_\mathrm{12th}$} & 
\makebox[1.5 cm][c]{$1-\alpha_\mathrm{tot}$} & 
\makebox[1.5 cm][c]{$\alpha_\mathrm{12th}$} & 
\makebox[1.5 cm][c]{$1-\alpha_\mathrm{tot}$} \\ \hline
 Q+V+W & 0.068\% & 99.0\% & 0.034\% & 99.5\% \\
 Q & 0.110\% & 98.4\% & - & - \\
 V & 0.024\% & $\geq$ 99.9\% & - & - \\
 W & 0.006\% & $\geq$ 99.9\%& - & - \\ 
\hline
\hline
\end{tabular}
\end{center}
\label{tb:alpha_tot}
\end{table}

The results we have obtained in this paper would pose a challenge
to the statistical isotropy and homogeneity hypothesis in cosmology, and
they should be examined through other independent tests. It will be
interesting to test in real space for scales corresponding to $220
\lesssim \ell \lesssim 240$. For example, a regional test of the power
spectrum was performed using a set of circular regions in real space
\cite{2004ApJ...605...14E}, and similar analyses were done using
different size and shape regions in \cite{2008JCAP...08..017L}. 
R\"{a}th {\it et al.} studied scale-dependent non-Gaussianities in a model-independent way 
and found significant anomalies in the multipole range $\ell = 120$-$300$
\cite{2009PhRvL.102m1301R,2011MNRAS.415.2205R,2011AdAst2011E..11R}.
Since we are intrigued with the scales at $220 \lesssim \ell \lesssim 240$, 
we should perform an analysis using finer segmentations.

Can inflationary cosmology explain the
anomalous feature found in this paper? Interesting discussions have been
done, for example, in \cite{2009PhRvD..80b3526D,2008PhRvD..78l3520E,2010PhRvD..81h3501C}. Although inflation tends to create homogeneous and
isotropic fluctuations in the mean and variance, they are on top of
the initially inhomogeneous and anisotropic background universe. Hence
the fluctuations could be statistically inhomogeneous and anisotropic in
total. Another reason would be that our universe is indeed anisotropic:
even if CMB fluctuations are statistically homogeneous and isotropic, the
CMB photons travel through the inhomogeneous universe, and the
gravitational lensing effects could impose statistically anisotropic
correlations \cite{2011arXiv1111.3357Rb}. In this case, the fluctuations are statistically
homogeneous and isotropic in ensemble average, and the anomaly we find
is a merely statistical fluctuation due to a particular realization of the
universe. It is worth studying how probable the observed inhomogeneous
universe can create an apparently anisotropic signal in the mean of the
CMB temperature fluctuations in harmonic space through the gravitational lensing effect.

%%%  SECTION 6 %%%
\section{Conclusion}
We have illustrated the method to test the statistical isotropy of the mean of the 
full sky CMB temperature fluctuations, which can be potentially useful for the forthcoming PLANCK data 
and future CMB surveys.

By applying the method to the WMAP seven-year observation data,  
We have found a
significant evidence for a non-zero mean of spherical harmonic
coefficients $a_{\ell m}$s at the multipole bin $\ell = 221$-$240$, 
for the combined Q+V+W map at the 99.93\% confidence level with the
KQ75y7 mask.  Using the more extensive mask, the
significance becomes larger, at the 99.96\% C.L.
As a whole, despite the 15 individual tests, this evidence against the
zero-mean hypothesis is still significant, keeping above the 99\% C.L.
This implies that we should challenge the zero-mean assumption and the
Universe could contain patterns on the characteristic scales.

However, since there might be unknown systematics, 
independent observations like PLANCK are awaited for further understanding. 

\section{Acknowledgments}
We would like to thank N. Sugiyama and S. Yokoyama for encouragements,
and B. Lew for his kind correspondences and useful
discussions at the early stage of this work.
Some of the results in this paper have bean obtained using the HEALPix
\cite{2005ApJ...622..759G} package. 
We acknowledge the use of
the Legacy Archive for Microwave Background Data Analysis (LAMBDA). 
Support for LAMBDA is provided by the NASA Office of Space Science.
This work has been also supported by the Grant-in-Aid for the Scientific
Research Fund (20740105, 23340046: TTT, 22012004: KI) commissioned by the
Ministry of Education, Culture, Sports, Science and Technology
(MEXT) of Japan.
We have been partially supported from the Grand-in-Aid for the
Global COE Program ``Quest for Fundamental Principles in the
Universe: from Particles to the Solar System and the Cosmos'' from
the MEXT.

\appendix
\section{contribution from the instrumental noise}
In order to assess the confidence levels of our findings we have performed Monte Carlo simulations
in which we took into account the anisotropic noise and the mask.
Based on the results of the simulations, we advocated anomalies in the means of multipole coefficients $a_{\ell m}$s.
Nevertheless you may think that the anomalies are due to the anisotropic noise.
We cannot rebut completely this suspicion because we will never know 
the exact information of  particular realization of the instrumental noise contained in the WMAP data.
However, it is possible to estimate the contribution from the anisotropic noise in the mean statistics by performing the same analysis for the CMB-free combination maps
(W1-W2, V1-V2, Q1-Q2)
, which are mainly determined by the instrumental noise.

\begin{figure}[htbp] %  figure placement: here, top, bottom, or page
   \centering
   \includegraphics[width = \linewidth]{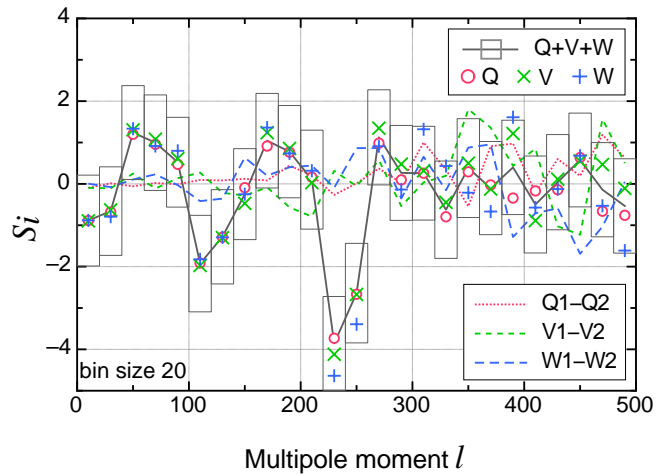} 
   \caption{The result from the CMB-free combination (dashed lines), over plotted on the lower panel in Fig. \ref{fig:band-mean-spec}. It seems that the contribution from the noise is smaller than that of the signal in the multipole range we are discussing and the behavior of the noise at $\ell=221$-$240$ does not show any anomaly. }
   \label{fig:cmbfree_binsize20}
\end{figure}

Figure \ref{fig:cmbfree_binsize20} shows the results, over plotted on the lower panel in Fig. \ref{fig:band-mean-spec}.
In order to see the effect of noise for higher multipole range,
we extend the range of multipole up to $\ell = 500$.
The values indicated by three dashed lines are computed as follows.
A CMB-free map is defined as 
\begin{equation}
T_{j1-j2} ( p) = T_{j1} ( p)-T_{j2}( p) ~,
\label{eq:cmbfree}
\end{equation}
where $p$ and $j$ are the pixel and DA indices respectively.
For this map $T_{j1-j2} (i)$,
we rewrite Eq. (\ref{eq:binned-mean-spec}) as
\begin{equation}
\mathcal{M}_i^{j1-j2} = \frac{\sum_{\ell=\ell_{-,i}}^{\ell_{+,i}} \sum_{m=-\ell}^{\ell} a^{j1-j2,\mathrm{mask}}_{\ell m} / \sqrt{\mathrm{Var}_\ell} }{\sqrt{2}\sqrt{\sum_{\ell =\ell_{-,i}}^{\ell_{+,i}} (2\ell+1)}}.
\label{eq:binned-mean-spec-cmbfree}
\end{equation}
Here $\mathrm{Var}_\ell$ is calculated from the three frequency band maps (Q, V, W) by Eq. (\ref{eq:Var_ell})
in order to compare with contribution from the instrumental noise with the signals.
The noise rms increases by a factor of $\sqrt{2}$ by executing Eq. (\ref{eq:cmbfree}) because the noises in different DA maps are uncorrelated.
Therefore Eq.(\ref{eq:binned-mean-spec-cmbfree}) is divid by $\sqrt{2}$.
Using $\mathcal{M}_i^{j1-j2}$ and the window matrix $W_{ij}$ calculated from the individual Q, V, W signal maps,
we construct decorrelated variables $S^{j1-j2}_i$ by using Eq. (\ref{eq:S_i^a}).
The values of $S_i^{j1-j2}$ represent noise-levels of a single map per DA.

In Fig. (\ref{fig:cmbfree_binsize20}), we see that  the contribution from
the noise is much smaller than that of the signal in the multipole range we are discussing and
also the behavior of the noise at the multipole bin $\ell = 221$-$240$ does not show any anomaly.
Further, in fact, expected noise-level of the combined maps should be lower than 
that of a single map per DA by a factor of $1/\sqrt{2}$ (Q, V-band) or $1/2$ (W-band)
because two or four maps are combined.
Therefore we may conclude that
our findings are not attributed to the anisotropic instrumental noise, but physically significant.

However, for multipole range $\ell \gtrsim 300$, clearly the noise becomes dominant
and the statistics behave independently across three frequency bands (Q, V, W).
Therefore the results in this paper ware obtained by using the multipole range up to $\ell = 300$.

\bibliography{refs}

%Merlin.mbs v4.21 2009-07-09.
\begin{thebibliography}{10}%
\makeatletter
\providecommand \@ifxundefined [1]{%
 \ifx #1\undefined \expandafter \@firstoftwo
 \else \expandafter \@secondoftwo
\fi
}%
\providecommand \@ifnum [1]{%
 \ifnum #1\expandafter \@firstoftwo
 \else \expandafter \@secondoftwo
\fi
}%
\providecommand \enquote [1]{``#1''}%
\providecommand \bibnamefont  [1]{#1}%
\providecommand \bibfnamefont [1]{#1}%
\providecommand \citenamefont [1]{#1}%
\providecommand\href[0]{\@sanitize\@href}%
\providecommand\@href[1]{\endgroup\@@startlink{#1}\endgroup\@@href}%
\providecommand\@@href[1]{#1\@@endlink}%
\providecommand \@sanitize [0]{\begingroup\catcode`\&12\catcode`\#12\relax}%
\@ifxundefined \pdfoutput {\@firstoftwo}{%
 \@ifnum{\z@=\pdfoutput}{\@firstoftwo}{\@secondoftwo}%
}{%
 \providecommand\@@startlink[1]{\leavevmode\special{html:<a href="#1">}}%
 \providecommand\@@endlink[0]{\special{html:</a>}}%
}{%
 \providecommand\@@startlink[1]{%
  \leavevmode
  \pdfstartlink
   attr{/Border[0 0 1 ]/H/I/C[0 1 1]}%
   user{/Subtype/Link/A<</Type/Action/S/URI/URI(#1)>>}%
  \relax
 }%
 \providecommand\@@endlink[0]{\pdfendlink}%
}%
\providecommand \url  [0]{\begingroup\@sanitize \@url }%
\providecommand \@url [1]{\endgroup\@href {#1}{\urlprefix}}%
\providecommand \urlprefix [0]{URL }%
\providecommand \Eprint[0]{\href }%
\@ifxundefined \urlstyle {%
  \providecommand \doi [1]{doi:\discretionary{}{}{}#1}%
}{%
  \providecommand \doi [0]{doi:\discretionary{}{}{}\begingroup
  \urlstyle{rm}\Url }%
}%
\providecommand \doibase [0]{http://dx.doi.org/}%
\providecommand \Doi[1]{\href{\doibase#1}}%
\providecommand \bibAnnote [3]{%
  \BibitemShut{#1}%
  \begin{quotation}\noindent
    \textsc{Key:}\ #2\\\textsc{Annotation:}\ #3%
  \end{quotation}%
}%
\providecommand \bibAnnoteFile [2]{%
  \IfFileExists{#2}{\bibAnnote {#1} {#2} {\input{#2}}}{}%
}%
\providecommand \typeout [0]{\immediate \write \m@ne }%
\providecommand \selectlanguage [0]{\@gobble}%
\providecommand \bibinfo [0]{\@secondoftwo}%
\providecommand \bibfield [0]{\@secondoftwo}%
\providecommand \translation [1]{[#1]}%
\providecommand \BibitemOpen[0]{}%
\providecommand \bibitemStop [0]{}%
\providecommand \bibitemNoStop [0]{.\EOS\space}%
\providecommand \EOS [0]{\spacefactor3000\relax}%
\providecommand \BibitemShut [1]{\csname bibitem#1\endcsname}%
%</preamble>
\bibitem{2005ppci.conf..235L}%
  \BibitemOpen
  \bibfield{author}{%
  \bibinfo {author} {\bibfnamefont{D.}~\bibnamefont{{Langlois}}},\ }%
  in\ \emph{\bibinfo {booktitle} {NATO ASIB Proc. 188: Particle Physics and
  Cosmology: the Interface}},\ \bibinfo {editor} {edited by\ \bibinfo {editor}
  {\bibnamefont{{D.~Kazakov \& G.~Smadja}}}}\ (\bibinfo {year} {2005})\ pp.\
  \bibinfo {pages} {235--+},\
  \Eprint{http://arxiv.org/abs/arXiv:hep-th/0405053}{arXiv:hep-th/0405053}%
  \bibAnnoteFile{NoStop}{2005ppci.conf..235L}%
\bibitem{2011CQGra..28p4003S}%
  \BibitemOpen
  \bibfield{author}{%
  \bibinfo {author} {\bibfnamefont{F.}~\bibnamefont{{Sylos Labini}}},\ }%
  \bibfield{journal}{%
  \Doi{10.1088/0264-9381/28/16/164003}{\bibinfo {journal} {Classical and
  Quantum Gravity}}\ }%
  \textbf{\bibinfo {volume} {28}},\ \bibinfo {pages} {164003} (\bibinfo {month}
  {Aug.}\ \bibinfo {year} {2011}),\
  \Eprint{http://arxiv.org/abs/1103.5974}{arXiv:1103.5974 [astro-ph.CO]}%
  \bibAnnoteFile{NoStop}{2011CQGra..28p4003S}%
\bibitem{2004MNRAS.349..313P}%
  \BibitemOpen
  \bibfield{author}{%
  \bibinfo {author} {\bibfnamefont{C.-G.}\ \bibnamefont{{Park}}},\ }%
  \bibfield{journal}{%
  \Doi{10.1111/j.1365-2966.2004.07500.x}{\bibinfo {journal} {MNRAS}}\ }%
  \textbf{\bibinfo {volume} {349}},\ \bibinfo {pages} {313} (\bibinfo {month}
  {Mar.}\ \bibinfo {year} {2004}),\
  \Eprint{http://arxiv.org/abs/arXiv:astro-ph/0307469}{arXiv:astro-ph/0307469}%
  \bibAnnoteFile{NoStop}{2004MNRAS.349..313P}%
\bibitem{2004ApJ...605...14E}%
  \BibitemOpen
  \bibfield{author}{%
  \bibinfo {author} {\bibfnamefont{H.~K.}\ \bibnamefont{{Eriksen}}}, \bibinfo
  {author} {\bibfnamefont{F.~K.}\ \bibnamefont{{Hansen}}}, \bibinfo {author}
  {\bibfnamefont{A.~J.}\ \bibnamefont{{Banday}}}, \bibinfo {author}
  {\bibfnamefont{K.~M.}\ \bibnamefont{{G{\'o}rski}}},\ and\ \bibinfo {author}
  {\bibfnamefont{P.~B.}\ \bibnamefont{{Lilje}}},\ }%
  \bibfield{journal}{%
  \Doi{10.1086/382267}{\bibinfo {journal} {\apj}}\ }%
  \textbf{\bibinfo {volume} {605}},\ \bibinfo {pages} {14} (\bibinfo {month}
  {Apr.}\ \bibinfo {year} {2004}),\
  \Eprint{http://arxiv.org/abs/arXiv:astro-ph/0307507}{arXiv:astro-ph/0307507}%
  \bibAnnoteFile{NoStop}{2004ApJ...605...14E}%
\bibitem{2004ApJ...609.1198E}%
  \BibitemOpen
  \bibfield{author}{%
  \bibinfo {author} {\bibfnamefont{H.~K.}\ \bibnamefont{{Eriksen}}}, \bibinfo
  {author} {\bibfnamefont{F.~K.}\ \bibnamefont{{Hansen}}}, \bibinfo {author}
  {\bibfnamefont{A.~J.}\ \bibnamefont{{Banday}}}, \bibinfo {author}
  {\bibfnamefont{K.~M.}\ \bibnamefont{{G{\'o}rski}}},\ and\ \bibinfo {author}
  {\bibfnamefont{P.~B.}\ \bibnamefont{{Lilje}}},\ }%
  \bibfield{journal}{%
  \Doi{10.1086/421972}{\bibinfo {journal} {\apj}}\ }%
  \textbf{\bibinfo {volume} {609}},\ \bibinfo {pages} {1198} (\bibinfo {month}
  {Jul.}\ \bibinfo {year} {2004})%
  \bibAnnoteFile{NoStop}{2004ApJ...609.1198E}%
\bibitem{2004MNRAS.354..641H}%
  \BibitemOpen
  \bibfield{author}{%
  \bibinfo {author} {\bibfnamefont{F.~K.}\ \bibnamefont{{Hansen}}}, \bibinfo
  {author} {\bibfnamefont{A.~J.}\ \bibnamefont{{Banday}}},\ and\ \bibinfo
  {author} {\bibfnamefont{K.~M.}\ \bibnamefont{{G{\'o}rski}}},\ }%
  \bibfield{journal}{%
  \Doi{10.1111/j.1365-2966.2004.08229.x}{\bibinfo {journal} {MNRAS}}\ }%
  \textbf{\bibinfo {volume} {354}},\ \bibinfo {pages} {641} (\bibinfo {month}
  {Nov.}\ \bibinfo {year} {2004}),\
  \Eprint{http://arxiv.org/abs/arXiv:astro-ph/0404206}{arXiv:astro-ph/0404206}%
  \bibAnnoteFile{NoStop}{2004MNRAS.354..641H}%
\bibitem{2008PhRvD..78f3531B}%
  \BibitemOpen
  \bibfield{author}{%
  \bibinfo {author} {\bibfnamefont{A.}~\bibnamefont{{Bernui}}},\ }%
  \bibfield{journal}{%
  \Doi{10.1103/PhysRevD.78.063531}{\bibinfo {journal} {\prd}}\ }%
  \textbf{\bibinfo {volume} {78}},\ \bibinfo {eid} {063531} (\bibinfo {month}
  {Sep.}\ \bibinfo {year} {2008}),\
  \Eprint{http://arxiv.org/abs/0809.0934}{arXiv:0809.0934}%
  \bibAnnoteFile{NoStop}{2008PhRvD..78f3531B}%
\bibitem{2009ApJ...704.1448H}%
  \BibitemOpen
  \bibfield{author}{%
  \bibinfo {author} {\bibfnamefont{F.~K.}\ \bibnamefont{{Hansen}}}, \bibinfo
  {author} {\bibfnamefont{A.~J.}\ \bibnamefont{{Banday}}}, \bibinfo {author}
  {\bibfnamefont{K.~M.}\ \bibnamefont{{G{\'o}rski}}}, \bibinfo {author}
  {\bibfnamefont{H.~K.}\ \bibnamefont{{Eriksen}}},\ and\ \bibinfo {author}
  {\bibfnamefont{P.~B.}\ \bibnamefont{{Lilje}}},\ }%
  \bibfield{journal}{%
  \Doi{10.1088/0004-637X/704/2/1448}{\bibinfo {journal} {\apj}}\ }%
  \textbf{\bibinfo {volume} {704}},\ \bibinfo {pages} {1448} (\bibinfo {month}
  {Oct.}\ \bibinfo {year} {2009}),\
  \Eprint{http://arxiv.org/abs/0812.3795}{arXiv:0812.3795}%
  \bibAnnoteFile{NoStop}{2009ApJ...704.1448H}%
\bibitem{2009ApJ...699..985H}%
  \BibitemOpen
  \bibfield{author}{%
  \bibinfo {author} {\bibfnamefont{J.}~\bibnamefont{{Hoftuft}}}, \bibinfo
  {author} {\bibfnamefont{H.~K.}\ \bibnamefont{{Eriksen}}}, \bibinfo {author}
  {\bibfnamefont{A.~J.}\ \bibnamefont{{Banday}}}, \bibinfo {author}
  {\bibfnamefont{K.~M.}\ \bibnamefont{{G{\'o}rski}}}, \bibinfo {author}
  {\bibfnamefont{F.~K.}\ \bibnamefont{{Hansen}}},\ and\ \bibinfo {author}
  {\bibfnamefont{P.~B.}\ \bibnamefont{{Lilje}}},\ }%
  \bibfield{journal}{%
  \Doi{10.1088/0004-637X/699/2/985}{\bibinfo {journal} {\apj}}\ }%
  \textbf{\bibinfo {volume} {699}},\ \bibinfo {pages} {985} (\bibinfo {month}
  {Jul.}\ \bibinfo {year} {2009}),\
  \Eprint{http://arxiv.org/abs/0903.1229}{arXiv:0903.1229 [astro-ph.CO]}%
  \bibAnnoteFile{NoStop}{2009ApJ...699..985H}%
\bibitem{2008JCAP...08..017L}%
  \BibitemOpen
  \bibfield{author}{%
  \bibinfo {author} {\bibfnamefont{B.}~\bibnamefont{{Lew}}},\ }%
  \bibfield{journal}{%
  \Doi{10.1088/1475-7516/2008/08/017}{\bibinfo {journal} {JCAP}}\ }%
  \textbf{\bibinfo {volume} {8}},\ \bibinfo {pages} {17} (\bibinfo {month}
  {Aug.}\ \bibinfo {year} {2008}),\
  \Eprint{http://arxiv.org/abs/0803.1409}{arXiv:0803.1409}%
  \bibAnnoteFile{NoStop}{2008JCAP...08..017L}%
\bibitem{2008PhRvD..78l3002N}%
  \BibitemOpen
  \bibfield{author}{%
  \bibinfo {author} {\bibfnamefont{R.}~\bibnamefont{{Nagata}}}\ and\ \bibinfo
  {author} {\bibfnamefont{J.}~\bibnamefont{{Yokoyama}}},\ }%
  \bibfield{journal}{%
  \Doi{10.1103/PhysRevD.78.123002}{\bibinfo {journal} {\prd}}\ }%
  \textbf{\bibinfo {volume} {78}},\ \bibinfo {pages} {123002} (\bibinfo {month}
  {Dec.}\ \bibinfo {year} {2008}),\
  \Eprint{http://arxiv.org/abs/0809.4537}{arXiv:0809.4537}%
  \bibAnnoteFile{NoStop}{2008PhRvD..78l3002N}%
\bibitem{2011JCAP...03..048A}%
  \BibitemOpen
  \bibfield{author}{%
  \bibinfo {author} {\bibfnamefont{C.}~\bibnamefont{{Armendariz-Picon}}},\ }%
  \bibfield{journal}{%
  \Doi{10.1088/1475-7516/2011/03/048}{\bibinfo {journal} {JCAP}}\ }%
  \textbf{\bibinfo {volume} {3}},\ \bibinfo {pages} {48} (\bibinfo {month}
  {Mar.}\ \bibinfo {year} {2011}),\
  \Eprint{http://arxiv.org/abs/1012.2849}{arXiv:1012.2849 [astro-ph.CO]}%
  \bibAnnoteFile{NoStop}{2011JCAP...03..048A}%
\bibitem{2011ApJS..192...15G}%
  \BibitemOpen
  \bibfield{author}{%
  \bibinfo {author} {\bibfnamefont{B.}~\bibnamefont{{Gold}}}, \bibinfo {author}
  {\bibfnamefont{N.}~\bibnamefont{{Odegard}}}, \bibinfo {author}
  {\bibfnamefont{J.~L.}\ \bibnamefont{{Weiland}}}, \bibinfo {author}
  {\bibfnamefont{R.~S.}\ \bibnamefont{{Hill}}}, \bibinfo {author}
  {\bibfnamefont{A.}~\bibnamefont{{Kogut}}}, \bibinfo {author}
  {\bibfnamefont{C.~L.}\ \bibnamefont{{Bennett}}}, \bibinfo {author}
  {\bibfnamefont{G.}~\bibnamefont{{Hinshaw}}}, \bibinfo {author}
  {\bibfnamefont{X.}~\bibnamefont{{Chen}}}, \bibinfo {author}
  {\bibfnamefont{J.}~\bibnamefont{{Dunkley}}}, \bibinfo {author}
  {\bibfnamefont{M.}~\bibnamefont{{Halpern}}}, \bibinfo {author}
  {\bibfnamefont{N.}~\bibnamefont{{Jarosik}}}, \bibinfo {author}
  {\bibfnamefont{E.}~\bibnamefont{{Komatsu}}}, \bibinfo {author}
  {\bibfnamefont{D.}~\bibnamefont{{Larson}}}, \bibinfo {author}
  {\bibfnamefont{M.}~\bibnamefont{{Limon}}}, \bibinfo {author}
  {\bibfnamefont{S.~S.}\ \bibnamefont{{Meyer}}}, \bibinfo {author}
  {\bibfnamefont{M.~R.}\ \bibnamefont{{Nolta}}}, \bibinfo {author}
  {\bibfnamefont{L.}~\bibnamefont{{Page}}}, \bibinfo {author}
  {\bibfnamefont{K.~M.}\ \bibnamefont{{Smith}}}, \bibinfo {author}
  {\bibfnamefont{D.~N.}\ \bibnamefont{{Spergel}}}, \bibinfo {author}
  {\bibfnamefont{G.~S.}\ \bibnamefont{{Tucker}}}, \bibinfo {author}
  {\bibfnamefont{E.}~\bibnamefont{{Wollack}}},\ and\ \bibinfo {author}
  {\bibfnamefont{E.~L.}\ \bibnamefont{{Wright}}},\ }%
  \bibfield{journal}{%
  \Doi{10.1088/0067-0049/192/2/15}{\bibinfo {journal} {ApJS}}\ }%
  \textbf{\bibinfo {volume} {192}},\ \bibinfo {pages} {15} (\bibinfo {month}
  {Feb.}\ \bibinfo {year} {2011}),\
  \Eprint{http://arxiv.org/abs/1001.4555}{arXiv:1001.4555 [astro-ph.GA]}%
  \bibAnnoteFile{NoStop}{2011ApJS..192...15G}%
\bibitem{2003ApJS..148...29J}%
  \BibitemOpen
  \bibfield{author}{%
  \bibinfo {author} {\bibfnamefont{N.}~\bibnamefont{{Jarosik}}}, \bibinfo
  {author} {\bibfnamefont{C.}~\bibnamefont{{Barnes}}}, \bibinfo {author}
  {\bibfnamefont{C.~L.}\ \bibnamefont{{Bennett}}}, \bibinfo {author}
  {\bibfnamefont{M.}~\bibnamefont{{Halpern}}}, \bibinfo {author}
  {\bibfnamefont{G.}~\bibnamefont{{Hinshaw}}}, \bibinfo {author}
  {\bibfnamefont{A.}~\bibnamefont{{Kogut}}}, \bibinfo {author}
  {\bibfnamefont{M.}~\bibnamefont{{Limon}}}, \bibinfo {author}
  {\bibfnamefont{S.~S.}\ \bibnamefont{{Meyer}}}, \bibinfo {author}
  {\bibfnamefont{L.}~\bibnamefont{{Page}}}, \bibinfo {author}
  {\bibfnamefont{D.~N.}\ \bibnamefont{{Spergel}}}, \bibinfo {author}
  {\bibfnamefont{G.~S.}\ \bibnamefont{{Tucker}}}, \bibinfo {author}
  {\bibfnamefont{J.~L.}\ \bibnamefont{{Weiland}}}, \bibinfo {author}
  {\bibfnamefont{E.}~\bibnamefont{{Wollack}}},\ and\ \bibinfo {author}
  {\bibfnamefont{E.~L.}\ \bibnamefont{{Wright}}},\ }%
  \bibfield{journal}{%
  \Doi{10.1086/377221}{\bibinfo {journal} {ApJS}}\ }%
  \textbf{\bibinfo {volume} {148}},\ \bibinfo {pages} {29} (\bibinfo {month}
  {Sep.}\ \bibinfo {year} {2003}),\
  \Eprint{http://arxiv.org/abs/arXiv:astro-ph/0302224}{arXiv:astro-ph/0302224}%
  \bibAnnoteFile{NoStop}{2003ApJS..148...29J}%
\bibitem{2007ApJS..170..263J}%
  \BibitemOpen
  \bibfield{author}{%
  \bibinfo {author} {\bibfnamefont{N.}~\bibnamefont{{Jarosik}}}, \bibinfo
  {author} {\bibfnamefont{C.}~\bibnamefont{{Barnes}}}, \bibinfo {author}
  {\bibfnamefont{M.~R.}\ \bibnamefont{{Greason}}}, \bibinfo {author}
  {\bibfnamefont{R.~S.}\ \bibnamefont{{Hill}}}, \bibinfo {author}
  {\bibfnamefont{M.~R.}\ \bibnamefont{{Nolta}}}, \bibinfo {author}
  {\bibfnamefont{N.}~\bibnamefont{{Odegard}}}, \bibinfo {author}
  {\bibfnamefont{J.~L.}\ \bibnamefont{{Weiland}}}, \bibinfo {author}
  {\bibfnamefont{R.}~\bibnamefont{{Bean}}}, \bibinfo {author}
  {\bibfnamefont{C.~L.}\ \bibnamefont{{Bennett}}}, \bibinfo {author}
  {\bibfnamefont{O.}~\bibnamefont{{Dor{\'e}}}}, \bibinfo {author}
  {\bibfnamefont{M.}~\bibnamefont{{Halpern}}}, \bibinfo {author}
  {\bibfnamefont{G.}~\bibnamefont{{Hinshaw}}}, \bibinfo {author}
  {\bibfnamefont{A.}~\bibnamefont{{Kogut}}}, \bibinfo {author}
  {\bibfnamefont{E.}~\bibnamefont{{Komatsu}}}, \bibinfo {author}
  {\bibfnamefont{M.}~\bibnamefont{{Limon}}}, \bibinfo {author}
  {\bibfnamefont{S.~S.}\ \bibnamefont{{Meyer}}}, \bibinfo {author}
  {\bibfnamefont{L.}~\bibnamefont{{Page}}}, \bibinfo {author}
  {\bibfnamefont{D.~N.}\ \bibnamefont{{Spergel}}}, \bibinfo {author}
  {\bibfnamefont{G.~S.}\ \bibnamefont{{Tucker}}}, \bibinfo {author}
  {\bibfnamefont{E.}~\bibnamefont{{Wollack}}},\ and\ \bibinfo {author}
  {\bibfnamefont{E.~L.}\ \bibnamefont{{Wright}}},\ }%
  \bibfield{journal}{%
  \Doi{10.1086/513697}{\bibinfo {journal} {ApJS}}\ }%
  \textbf{\bibinfo {volume} {170}},\ \bibinfo {pages} {263} (\bibinfo {month}
  {Jun.}\ \bibinfo {year} {2007}),\
  \Eprint{http://arxiv.org/abs/arXiv:astro-ph/0603452}{arXiv:astro-ph/0603452}%
  \bibAnnoteFile{NoStop}{2007ApJS..170..263J}%
\bibitem{2011ApJS..192...14J}%
  \BibitemOpen
  \bibfield{author}{%
  \bibinfo {author} {\bibfnamefont{N.}~\bibnamefont{{Jarosik}}}, \bibinfo
  {author} {\bibfnamefont{C.~L.}\ \bibnamefont{{Bennett}}}, \bibinfo {author}
  {\bibfnamefont{J.}~\bibnamefont{{Dunkley}}}, \bibinfo {author}
  {\bibfnamefont{B.}~\bibnamefont{{Gold}}}, \bibinfo {author}
  {\bibfnamefont{M.~R.}\ \bibnamefont{{Greason}}}, \bibinfo {author}
  {\bibfnamefont{M.}~\bibnamefont{{Halpern}}}, \bibinfo {author}
  {\bibfnamefont{R.~S.}\ \bibnamefont{{Hill}}}, \bibinfo {author}
  {\bibfnamefont{G.}~\bibnamefont{{Hinshaw}}}, \bibinfo {author}
  {\bibfnamefont{A.}~\bibnamefont{{Kogut}}}, \bibinfo {author}
  {\bibfnamefont{E.}~\bibnamefont{{Komatsu}}}, \bibinfo {author}
  {\bibfnamefont{D.}~\bibnamefont{{Larson}}}, \bibinfo {author}
  {\bibfnamefont{M.}~\bibnamefont{{Limon}}}, \bibinfo {author}
  {\bibfnamefont{S.~S.}\ \bibnamefont{{Meyer}}}, \bibinfo {author}
  {\bibfnamefont{M.~R.}\ \bibnamefont{{Nolta}}}, \bibinfo {author}
  {\bibfnamefont{N.}~\bibnamefont{{Odegard}}}, \bibinfo {author}
  {\bibfnamefont{L.}~\bibnamefont{{Page}}}, \bibinfo {author}
  {\bibfnamefont{K.~M.}\ \bibnamefont{{Smith}}}, \bibinfo {author}
  {\bibfnamefont{D.~N.}\ \bibnamefont{{Spergel}}}, \bibinfo {author}
  {\bibfnamefont{G.~S.}\ \bibnamefont{{Tucker}}}, \bibinfo {author}
  {\bibfnamefont{J.~L.}\ \bibnamefont{{Weiland}}}, \bibinfo {author}
  {\bibfnamefont{E.}~\bibnamefont{{Wollack}}},\ and\ \bibinfo {author}
  {\bibfnamefont{E.~L.}\ \bibnamefont{{Wright}}},\ }%
  \bibfield{journal}{%
  \Doi{10.1088/0067-0049/192/2/14}{\bibinfo {journal} {ApJS}}\ }%
  \textbf{\bibinfo {volume} {192}},\ \bibinfo {pages} {14} (\bibinfo {month}
  {Feb.}\ \bibinfo {year} {2011}),\
  \Eprint{http://arxiv.org/abs/1001.4744}{arXiv:1001.4744 [astro-ph.CO]}%
  \bibAnnoteFile{NoStop}{2011ApJS..192...14J}%
\bibitem{2005ApJ...622..759G}%
  \BibitemOpen
  \bibfield{author}{%
  \bibinfo {author} {\bibfnamefont{K.~M.}\ \bibnamefont{{G{\'o}rski}}},
  \bibinfo {author} {\bibfnamefont{E.}~\bibnamefont{{Hivon}}}, \bibinfo
  {author} {\bibfnamefont{A.~J.}\ \bibnamefont{{Banday}}}, \bibinfo {author}
  {\bibfnamefont{B.~D.}\ \bibnamefont{{Wandelt}}}, \bibinfo {author}
  {\bibfnamefont{F.~K.}\ \bibnamefont{{Hansen}}}, \bibinfo {author}
  {\bibfnamefont{M.}~\bibnamefont{{Reinecke}}},\ and\ \bibinfo {author}
  {\bibfnamefont{M.}~\bibnamefont{{Bartelmann}}},\ }%
  \bibfield{journal}{%
  \Doi{10.1086/427976}{\bibinfo {journal} {\apj}}\ }%
  \textbf{\bibinfo {volume} {622}},\ \bibinfo {pages} {759} (\bibinfo {month}
  {Apr.}\ \bibinfo {year} {2005}),\
  \Eprint{http://arxiv.org/abs/arXiv:astro-ph/0409513}{arXiv:astro-ph/0409513}%
  \bibAnnoteFile{NoStop}{2005ApJ...622..759G}%
\bibitem{LAMBDAwebsite}%
  \BibitemOpen
  \bibfield{author}{%
  \bibinfo {author} {\bibnamefont{{LAMBDA website}}},\ }%
  \bibinfo {note} {\url{http://lambda.gsfc.nasa.gov/}}%
  \bibAnnoteFile{NoStop}{LAMBDAwebsite}%
\bibitem{2011ApJS..192...16L}%
  \BibitemOpen
  \bibfield{author}{%
  \bibinfo {author} {\bibfnamefont{D.}~\bibnamefont{{Larson}}}, \bibinfo
  {author} {\bibfnamefont{J.}~\bibnamefont{{Dunkley}}}, \bibinfo {author}
  {\bibfnamefont{G.}~\bibnamefont{{Hinshaw}}}, \bibinfo {author}
  {\bibfnamefont{E.}~\bibnamefont{{Komatsu}}}, \bibinfo {author}
  {\bibfnamefont{M.~R.}\ \bibnamefont{{Nolta}}}, \bibinfo {author}
  {\bibfnamefont{C.~L.}\ \bibnamefont{{Bennett}}}, \bibinfo {author}
  {\bibfnamefont{B.}~\bibnamefont{{Gold}}}, \bibinfo {author}
  {\bibfnamefont{M.}~\bibnamefont{{Halpern}}}, \bibinfo {author}
  {\bibfnamefont{R.~S.}\ \bibnamefont{{Hill}}}, \bibinfo {author}
  {\bibfnamefont{N.}~\bibnamefont{{Jarosik}}}, \bibinfo {author}
  {\bibfnamefont{A.}~\bibnamefont{{Kogut}}}, \bibinfo {author}
  {\bibfnamefont{M.}~\bibnamefont{{Limon}}}, \bibinfo {author}
  {\bibfnamefont{S.~S.}\ \bibnamefont{{Meyer}}}, \bibinfo {author}
  {\bibfnamefont{N.}~\bibnamefont{{Odegard}}}, \bibinfo {author}
  {\bibfnamefont{L.}~\bibnamefont{{Page}}}, \bibinfo {author}
  {\bibfnamefont{K.~M.}\ \bibnamefont{{Smith}}}, \bibinfo {author}
  {\bibfnamefont{D.~N.}\ \bibnamefont{{Spergel}}}, \bibinfo {author}
  {\bibfnamefont{G.~S.}\ \bibnamefont{{Tucker}}}, \bibinfo {author}
  {\bibfnamefont{J.~L.}\ \bibnamefont{{Weiland}}}, \bibinfo {author}
  {\bibfnamefont{E.}~\bibnamefont{{Wollack}}},\ and\ \bibinfo {author}
  {\bibfnamefont{E.~L.}\ \bibnamefont{{Wright}}},\ }%
  \bibfield{journal}{%
  \Doi{10.1088/0067-0049/192/2/16}{\bibinfo {journal} {ApJS}}\ }%
  \textbf{\bibinfo {volume} {192}},\ \bibinfo {pages} {16} (\bibinfo {month}
  {Feb.}\ \bibinfo {year} {2011}),\
  \Eprint{http://arxiv.org/abs/1001.4635}{arXiv:1001.4635 [astro-ph.CO]}%
  \bibAnnoteFile{NoStop}{2011ApJS..192...16L}%
\bibitem{2009PhRvL.102m1301R}%
  \BibitemOpen
  \bibfield{author}{%
  \bibinfo {author} {\bibfnamefont{C.}~\bibnamefont{{R{\"a}th}}}, \bibinfo
  {author} {\bibfnamefont{G.~E.}\ \bibnamefont{{Morfill}}}, \bibinfo {author}
  {\bibfnamefont{G.}~\bibnamefont{{Rossmanith}}}, \bibinfo {author}
  {\bibfnamefont{A.~J.}\ \bibnamefont{{Banday}}},\ and\ \bibinfo {author}
  {\bibfnamefont{K.~M.}\ \bibnamefont{{G{\'o}rski}}},\ }%
  \bibfield{journal}{%
  \Doi{10.1103/PhysRevLett.102.131301}{\bibinfo {journal} {Physical Review
  Letters}}\ }%
  \textbf{\bibinfo {volume} {102}},\ \bibinfo {eid} {131301} (\bibinfo {month}
  {Apr.}\ \bibinfo {year} {2009}),\
  \Eprint{http://arxiv.org/abs/0810.3805}{arXiv:0810.3805}%
  \bibAnnoteFile{NoStop}{2009PhRvL.102m1301R}%
\bibitem{2011MNRAS.415.2205R}%
  \BibitemOpen
  \bibfield{author}{%
  \bibinfo {author} {\bibfnamefont{C.}~\bibnamefont{{R{\"a}th}}}, \bibinfo
  {author} {\bibfnamefont{A.~J.}\ \bibnamefont{{Banday}}}, \bibinfo {author}
  {\bibfnamefont{G.}~\bibnamefont{{Rossmanith}}}, \bibinfo {author}
  {\bibfnamefont{H.}~\bibnamefont{{Modest}}}, \bibinfo {author}
  {\bibfnamefont{R.}~\bibnamefont{{S{\"u}tterlin}}}, \bibinfo {author}
  {\bibfnamefont{K.~M.}\ \bibnamefont{{G{\'o}rski}}}, \bibinfo {author}
  {\bibfnamefont{J.}~\bibnamefont{{Delabrouille}}},\ and\ \bibinfo {author}
  {\bibfnamefont{G.~E.}\ \bibnamefont{{Morfill}}},\ }%
  \bibfield{journal}{%
  \Doi{10.1111/j.1365-2966.2011.18844.x}{\bibinfo {journal} {MNRAS}}\ }%
  \textbf{\bibinfo {volume} {415}},\ \bibinfo {pages} {2205} (\bibinfo {month}
  {Aug.}\ \bibinfo {year} {2011}),\
  \Eprint{http://arxiv.org/abs/1012.2985}{arXiv:1012.2985 [astro-ph.CO]}%
  \bibAnnoteFile{NoStop}{2011MNRAS.415.2205R}%
\bibitem{2011AdAst2011E..11R}%
  \BibitemOpen
  \bibfield{author}{%
  \bibinfo {author} {\bibfnamefont{G.}~\bibnamefont{{Rossmanith}}}, \bibinfo
  {author} {\bibfnamefont{H.}~\bibnamefont{{Modest}}}, \bibinfo {author}
  {\bibfnamefont{C.}~\bibnamefont{{R{\"a}th}}}, \bibinfo {author}
  {\bibfnamefont{A.~J.}\ \bibnamefont{{Banday}}}, \bibinfo {author}
  {\bibfnamefont{K.~M.}\ \bibnamefont{{G{\'o}rski}}},\ and\ \bibinfo {author}
  {\bibfnamefont{G.}~\bibnamefont{{Morfill}}},\ }%
  \bibfield{journal}{%
  \bibinfo {journal} {Advances in Astronomy}\ }%
  \textbf{\bibinfo {volume} {2011}},\ \bibinfo {eid} {174873} (\bibinfo {year}
  {2011}),\ \doi{\bibinfo {doi} {10.1155/2011/174873}},\
  \Eprint{http://arxiv.org/abs/1108.0596}{arXiv:1108.0596 [astro-ph.CO]}%
  \bibAnnoteFile{NoStop}{2011AdAst2011E..11R}%
\bibitem{2009PhRvD..80b3526D}%
  \BibitemOpen
  \bibfield{author}{%
  \bibinfo {author} {\bibfnamefont{J.~F.}\ \bibnamefont{{Donoghue}}}, \bibinfo
  {author} {\bibfnamefont{K.}~\bibnamefont{{Dutta}}},\ and\ \bibinfo {author}
  {\bibfnamefont{A.}~\bibnamefont{{Ross}}},\ }%
  \bibfield{journal}{%
  \Doi{10.1103/PhysRevD.80.023526}{\bibinfo {journal} {\prd}}\ }%
  \textbf{\bibinfo {volume} {80}},\ \bibinfo {eid} {023526} (\bibinfo {month}
  {Jul.}\ \bibinfo {year} {2009}),\
  \Eprint{http://arxiv.org/abs/arXiv:astro-ph/0703455}{arXiv:astro-ph/0703455}%
  \bibAnnoteFile{NoStop}{2009PhRvD..80b3526D}%
\bibitem{2008PhRvD..78l3520E}%
  \BibitemOpen
  \bibfield{author}{%
  \bibinfo {author} {\bibfnamefont{A.~L.}\ \bibnamefont{{Erickcek}}}, \bibinfo
  {author} {\bibfnamefont{M.}~\bibnamefont{{Kamionkowski}}},\ and\ \bibinfo
  {author} {\bibfnamefont{S.~M.}\ \bibnamefont{{Carroll}}},\ }%
  \bibfield{journal}{%
  \Doi{10.1103/PhysRevD.78.123520}{\bibinfo {journal} {\prd}}\ }%
  \textbf{\bibinfo {volume} {78}},\ \bibinfo {eid} {123520} (\bibinfo {month}
  {Dec.}\ \bibinfo {year} {2008}),\
  \Eprint{http://arxiv.org/abs/0806.0377}{arXiv:0806.0377}%
  \bibAnnoteFile{NoStop}{2008PhRvD..78l3520E}%
\bibitem{2010PhRvD..81h3501C}%
  \BibitemOpen
  \bibfield{author}{%
  \bibinfo {author} {\bibfnamefont{S.~M.}\ \bibnamefont{{Carroll}}}, \bibinfo
  {author} {\bibfnamefont{C.-Y.}\ \bibnamefont{{Tseng}}},\ and\ \bibinfo
  {author} {\bibfnamefont{M.~B.}\ \bibnamefont{{Wise}}},\ }%
  \bibfield{journal}{%
  \Doi{10.1103/PhysRevD.81.083501}{\bibinfo {journal} {\prd}}\ }%
  \textbf{\bibinfo {volume} {81}},\ \bibinfo {eid} {083501} (\bibinfo {month}
  {Apr.}\ \bibinfo {year} {2010}),\
  \Eprint{http://arxiv.org/abs/0811.1086}{arXiv:0811.1086}%
  \bibAnnoteFile{NoStop}{2010PhRvD..81h3501C}%
\bibitem{2011arXiv1111.3357Rb}%
  \BibitemOpen
  \bibfield{author}{%
  \bibinfo {author} {\bibfnamefont{A.}~\bibnamefont{{Rotti}}}, \bibinfo
  {author} {\bibfnamefont{M.}~\bibnamefont{{Aich}}},\ and\ \bibinfo {author}
  {\bibfnamefont{T.}~\bibnamefont{{Souradeep}}},\ }%
  \bibfield{journal}{%
  \bibinfo {journal} {ArXiv e-prints}}%
   (\bibinfo {month} {Nov.}\ \bibinfo {year} {2011}),\
  \Eprint{http://arxiv.org/abs/1111.3357}{arXiv:1111.3357 [astro-ph.CO]}%
  \bibAnnoteFile{NoStop}{2011arXiv1111.3357Rb}%
\end{thebibliography}%

\end{document}